# Unibody microscope objective tipped with a microsphere: design, fabrication and application in subwavelength imaging

BING YAN,[1,2] YANG SONG,[2] XIBIN YANG,[2,3] DAXI XIONG,[2] AND ZENGBO WANG[1,*]

[1] *School of Computer Science and Electronic Engineering, Bangor University, Dean Street, Bangor, Gwynedd, LL57 1UT, UK*
[2] *Center of Optics Health, Suzhou Institute of Biomedical Engineering and Technology, Chinese Academy of Sciences, No. 88 Keling Street, Suzhou Jiangsu, 215163, China*
[3] *yangxb@sibet.ac.cn*
*\* z.wang@bangor.ac.uk*

**Abstract:** Microsphere-based subwavelength imaging technique was first demonstrated in 2011. After nearly a decade of efforts, such technique has spawned numerous interests in fields such as laser nano-machining, imaging, sensing and biological detection. For wider industrial scale application of the technique, a robust, low-cost objective lens incorporating microsphere lens is highly desired and sought by many researchers. In this work, we demonstrate a unibody microscope objective lens formed by tipping a high-index microsphere onto a Plano-Convex lens and subsequently fitting them into a conventional objective lens. We call such new objective the Plano-Convex-Microsphere (PCM) objective, which resembles the appearance and operation of an ordinary microscope objective, while providing super-resolving power in discerning subwavelength nanoscale features in air condition. The imaging performance of PCM-objective along with the working distance has been systematically investigated. With the assistance of scanning process, larger area imaging is realized. The PCM-objective can be easily adaptive to many existing microscope systems and appealing for commercialization.

## 1. Introduction

With advancement of life science and nano-technology, research has urgent demands for novel instruments to achieve nano-scale observation, which also prompt a revolution of optical microscopy. One of the great challenges in optical imaging is to break the diffraction limit, which was formulated as $d=\lambda/2NA$ (d is minimum resolvable distance; $\lambda$ is the wavelength of light; NA is the numerical aperture of the objective) [1]. In 1984, the development of near-field scanning optical microscopy (NSOM), which introduced the near-field method to collect evanescent waves by a physical probe with subwavelength aperture positioned close to an object's surface, that opens a new door for super-resolution research [2]. In recent years, owing to the rise of metamaterials, nanophotonics, and plasmonics, we witnessed a number of exciting developments in super-resolution imaging, including, for example, super-resolution fluorescent microscopy [3-5], negative-index metamaterial superlens [6,7], super-oscillation lens [8], time-reversal imaging [9], the Maxwell fisheye [10] and the scattering lens [11]. However, none of them can operate under white-light illuminations. Recently, it was discovered that dielectric spherical lens with micro-scale diameter can surpass the diffraction barrier through a phenomenon known as "photonic nanojet" [12,13]. This has been proven to be a simple and superior way to achieve a resolution of $\sim6/\lambda$ to $\sim\lambda/8$ in white-light conditions [14,15]. The super-resolution capability could be explained by efficient collection and conversion of the high spatial harmonics of evanescent waves of underlying object in near-field zone. Unlike other super-resolution techniques, microsphere superlens provides a real-time visualization under white-light illumination, meanwhile, it is label-free and low intrinsic loss at higher optical frequency. These unique features have attracted a number of groups across the world for further development. For instance, incorporation of high-index microsphere into flexible elastomers was reported to improve its durability and reusability for observations at a specific position [16,17]. Microsphere was coupled with a confocal microscope to improve the maximal resolvability to sub-50nm scale [18,19,20]. Furthermore, micro-lens nanoscopes in extraordinary forms were also demonstrated, such as nanoparticle-based metamaterial hemisphere superlens leading to white-lighting 45 nm resolution [21], and biological superlens using spider silks and cells [22, 23].

Although literatures revealed the capability of super-resolution for microsphere-based superlens, the imaging field of view (FOV) is often very small, typically only a quarter of microsphere's diameter, which is an inevitable issue in practical use. Efforts have been made recently to circumvent such problem. Our previous work presented in 2017 provided a simple but efficient approach to adopt microsphere lens in an ordinary microscope. With a 3D printed objective adaptor, the microsphere can work synchronously under objective to realize large-area scanning imaging [24]. Similar work was also demonstrated from H. Yang's previous paper [25]. However, due to lack of feedback mechanisms, these lenses still make contact with samples or require lubricants to reduce the friction during scanning, which adds invasiveness to sensitive specimens. In addition, optical trapping method was reported to control the microsphere, however, this method was limited to liquid environment [26]. The most effective way among the previously published research is the manipulation of single microsphere lens by tip-based scanning techniques [27-29]. In 2016, F. Wang et al. bonded BTG microsphere onto an AFM tip, with the advantage of precise positioning and feedback monitoring, the system has high precision in maintaining distance between microsphere and objects. Therefore, samples possessing different roughness surfaces can be either super-resolved by contact or constant-working-distance scanning mode [28]. Nevertheless, this technique involves sophisticated and pricey near-field tip control system, which requires professional skill to operate. Also, their work has been limited to use in liquid environment. We wish to develop a low-cost, ease-to-implement and universal microsphere-based super-resolution scanning imaging system for non-invasive nano-imaging, suitable for both air and liquid environment.

In this work, we employed a new design where the microsphere lens is a compound part consisting of a Plano-Convex lens and a Microsphere lens (in short "PCM"), which can perform as individual optical probe and is able to be perfectly fitted onto an ordinary objective lens to form a unibody PCM-objective lens. The unibody superlens has the same appearance and simple operation as ordinary microscope objective, but also providing non-invasive real-time super-resolving capability in air condition. Its imaging performance was evaluated along with the working distance. Meanwhile, with assist of high-resolution nano-stage, large-area scanning imaging was realized. Furthermore, the focusing properties of PCM lens was theoretically investigated. This unibody PCM-objective design is highly versatile, meaning that existing optical microscope system is able to be converted to super-resolution nanoscope by simply installing our lens. The work has raised the usability of microsphere-based superlens technique to a new level.

## 2. Preparation of unibody PCM-objective

In this paper, a unibody PCM-objective is presented in the form of combination of Plano-convex microsphere lens and an ordinary microscope objective. As previously reported, the PCM lens with advantage of excellent reliability and usability can generate sub-diffraction photonic nanojet in air condition, which is a low-cost and simple technique used for laser nano-patterning [30]. Here, we introduce the PCM lens to imaging application through integrating it onto an optical microscope objective. As Fig. 1(a) illustrating, a barium titanium glass (BTG) microsphere ($n_p$=1.9) with diameter of 25μm, 38μm and 50 μm were respectively aligned and attached onto polydimethylsiloxane (PDMS, SYLGARD 184, DOW CORNING, $n_m$=1.4) pre-coated (spin-coating at 2000 rpm for 1 minute) spherical crown of a BK7 Plano-Convex lens (LA1700, Thorlabs). Due to the capillary force of viscous PDMS liquid, once BTG microsphere was briefly touched the edge of PDMS and it was "sucked" and immersed into PDMS layer to form partially encapsulation of BTG microsphere. Then the PDMS was solidified through 90℃ 15minutes baking. This results in the formation of a probe-like Plano-Convex microsphere (PCM) lens, where a single microsphere slightly extruding out. Afterwards, the PCM lens was attached onto a custom-designed adjustable objective adapter. The adapter consists of two metal components, the upper part is fixed to the objective (40× 0.6NA) and lower part holds the PCM lens. These two components are fitted together with fine screw thread, hence the distance between PCM lens and objective is adjustable by manual rotation. Then, the PCM-objective was installed onto self-built imaging system as Fig. 1(b) shows. A blue LED light source (central wavelength 470nm, bandwidth 24 nm) incidents through the Kohler arrangement with an aperture diaphragm (AD) and field diaphragm (FD) which can contribute to achieving better image contrast. In our experiments, the PCM-objective was kept static and the underlying stage moved and scanned the samples across the lens. The movement of sample was performed using a high-resolution nano-stage (Nanomotor SNM01, Shinopto), with 20 nm resolution in the XYZ direction, and a travel range of 25.4 mm. In order to avoid damage of PCM lens and sample during focus adjustment, a long focal length zoom lens was placed horizontally at side for monitoring the gap between microsphere bottom and sample. The exact working distance (WD) can be known by measuring the travel distance of the nano-stage from working plane towards lens boundary, once the sample touching the lens, a slight pressure can cause deformation of image, which determines the contact state.

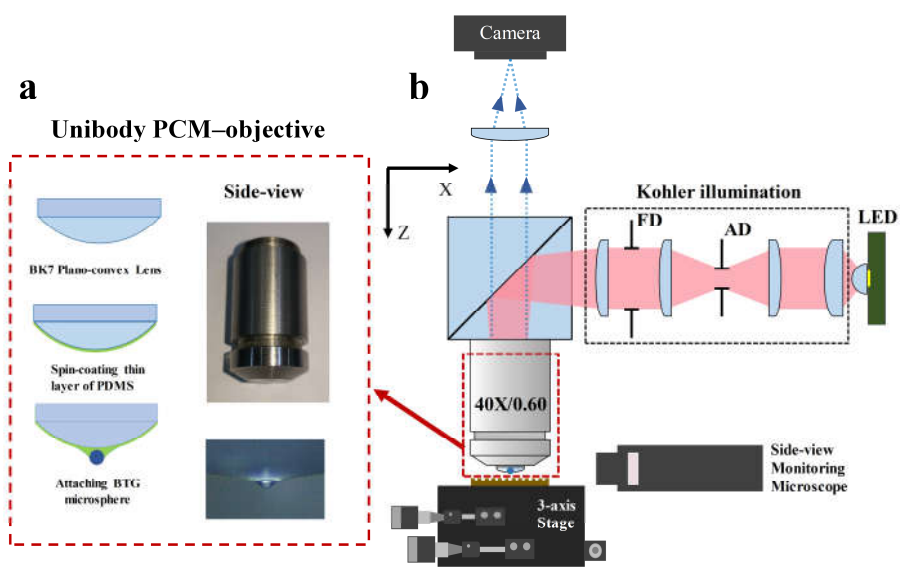

Fig. 1. Schematic of unibody PCM-objective imaging system. (a) Preparation of PCM lens and installation of unibody PCM-objective. (b) Self-developed imaging system.

## 3. Results and discussions

*3.1 Investigation of imaging performance*

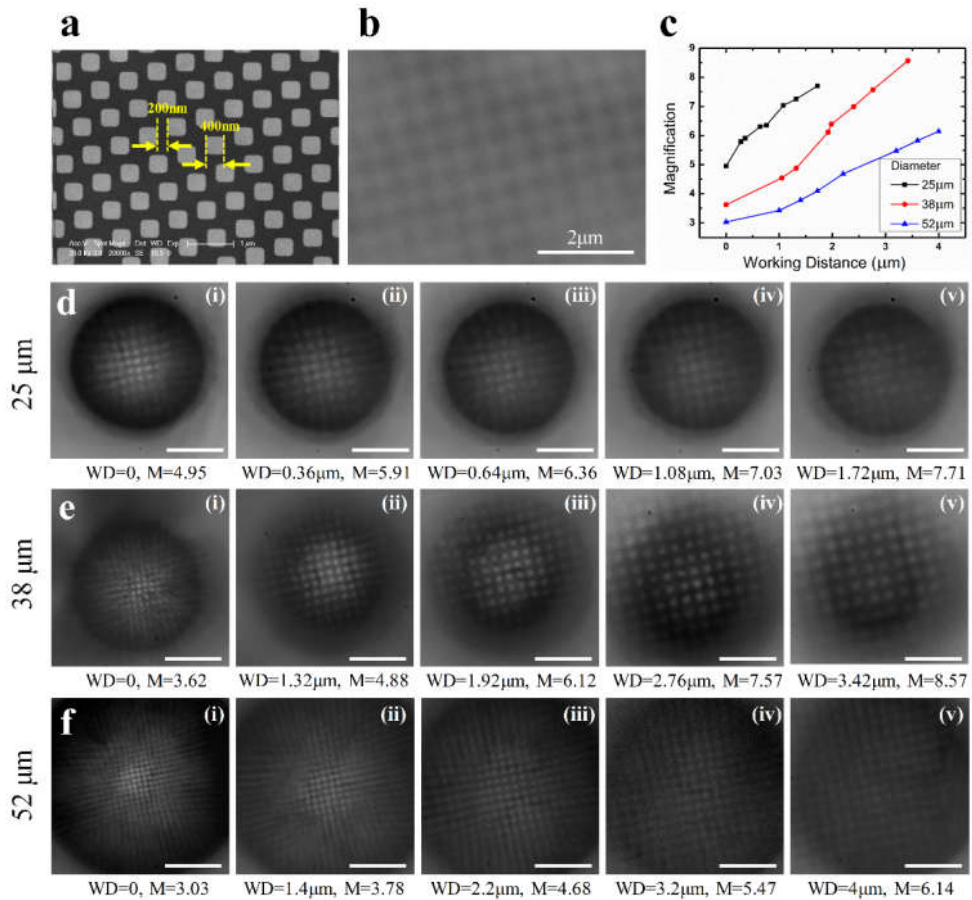

Fig. 2. Investigation of magnifications (M) of PCM lens from different working distances (WD). (a) SEM image of IC chip with 400nm blocks and 200nm intervals. (b) IC chip image taken by optical objective (40 X, 0.6 N.A.). (c) The curve of magnification of three sizes of microspheres with working distances. (d)(e)(f) are PCM lens with 25μm, 38μm and 52μm BTG microsphere embedded. Magnifications (M) of each size lens was evaluated by different WD. Scale bars in (b) is 2μm and in (d)(e)(f) are 20 μm.

Near-field imaging requires working distance within wavelength scale or even in contact manner. However, it is invasive and may bring limitations in sensitive specimens imaging, for instance biological samples. Therefore, contactless process is essential to avoid polluting and damaging specimens. In experiments, we evaluated the imaging performance including magnification and resolvability of developed unibody PCM-objective at different WD. The experiment was carried out by using a sample of IC chip with features of 400nm blocks and 200nm intervals, as showing in Fig. 2(a). Fig. 2(b) indicates the image obtained using a bare objective (40× NA0.60), which has theoretical resolution limit of 390nm (λ/2NA). As the minimum feature size is beyond the resolvability of the objective, it gave a blurry image and detail information was lost. In order to evaluate the imaging enhancement of PCM lens, we kept all conditions (objective, illumination and camera) the same and only installed PCM lenses of different diameters. Since the sample was placed within the focal length of bottom PCM lens, the top objective picked up the magnified virtual image generated by the PCM lens. Fig. 2(d-f) illustrate the magnifications of the images obtained by the microsphere of 25μm, 38μm and 52μm when the WD increases from zero, respectively. As can be seen, features of 200 nm separations can be clearly resolved through the contribution of PCM lens, meanwhile, the magnification ascents along with distance increasing, which follows geometrical optics. That is to say, the PCM lens not only has feature of classical lens, but also contributes significantly to enhance the resolution. Furthermore, comparison between three sizes of microspheres also reveals that PCM lens with smaller microsphere exhibits greater magnification factor at same working distance, which is summarized in Fig. 2(c). It is worth noting that the imaging pincushion distortion appeared apparently in case of lens adjacent to object and this phenomenon was gradually suppressed with distance expanding. For example, in case of contact (WD=0), it is clear to tell that the small area in the central region has better resolution and contrast than the periphery of microsphere, concurrently a radially increment of magnification factor resulting in distorted image, as shown in Fig. 2[d(i), e(i), f(i)]. This heterogeneous imaging performance may be due to the aberration of geometric construction of spherical lens. With lens moving away from the object, the uneven imaging performance and distortion is gradually suppressed, but the contrast and FOV are reduced correspondently. Therefore, there is a compromise between image quality and magnification factor and it is necessary to make a trade-off to get most balanced performance.

Besides, smaller size sample were used to further investigate super-resolution ability of designed unibody PCM-objective. Here, we used 38 μm microsphere PCM lens to capture Blu-ray disc with features of 200 nm stripes and 100 nm grooves [Fig. 3(a)] in different WD. It may be hard to resolve these patterns by standard illumination setting. However, using oblique illumination with an incident angle roughly between 10 and 40 degrees and partial illumination would produce an enhancement in image quality. This was realized by adjustment of aperture diaphragm (AD) and field diaphragm (FD). These illumination conditions for improving imaging performance were also discovered and proved experimentally and theoretically by our previous paper and other group's works [20,22,26]. According to results demonstrated and discussed in Fig. 2(e), the imaging performance is better balanced with WD range of 1μm to 3.5 μm. Therefore, Blu-ray disc imaging experiment was carried out within a corresponding WD range. As can be seen at 1.3 μm WD [Fig. 3(b)], imaging zone under microsphere is partially illuminated with a shadow at one size. Although 100 nm feature can be resolved, the center of illuminated region was nearly overexposed which affected image quality. Due to inclined and partial illumination, the brightness distribution of the image is uneven, especially in the shorter WD. This high brightness difference was gradually improved with WD increased. At WD around 3 μm [Fig. 3(e, f)], the image exposure became more equalized. The feature of Blu-ray disc can be clearly distinguished with magnification factor of ~8. After 4 μm, images turned into misty and lost details [Fig. 3(h, i)]. It was observed that the WD of ~4 μm is a critical experimental value, above which super-resolution is no longer possible.

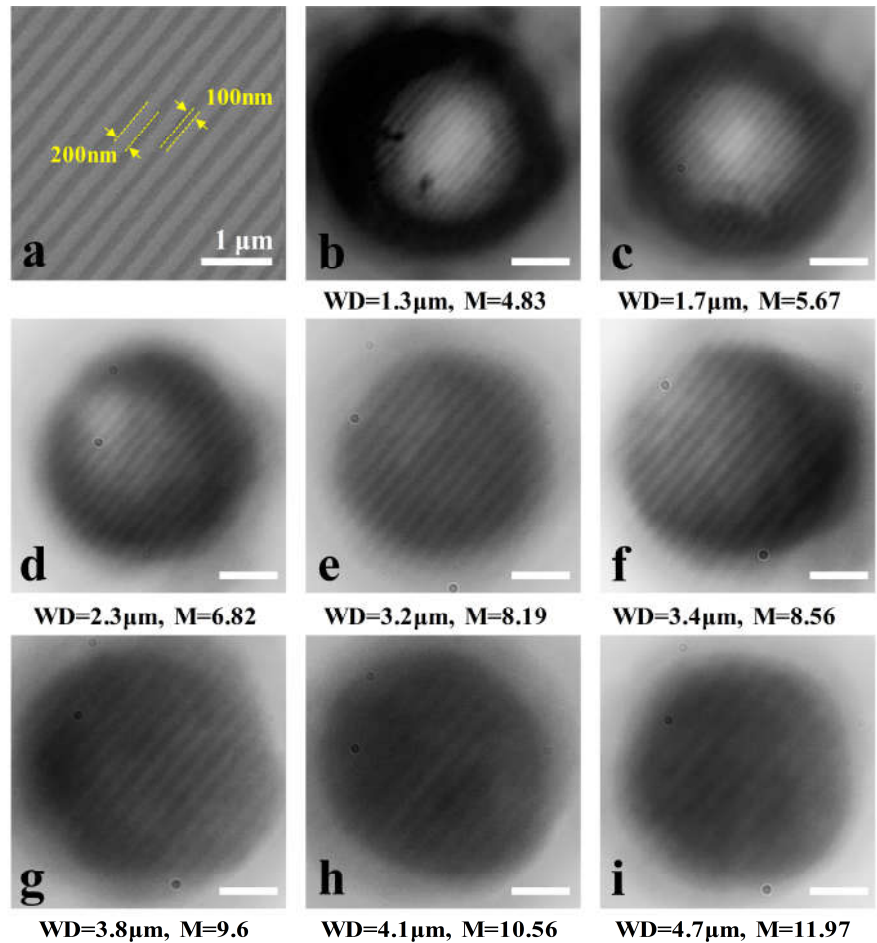

Fig. 3. Investigation of resolvability of PCM lens from different working distances (WD). (a) SEM image of Blu-ray disc with 200nm strips and 100nm grooves. (b)-(i) Blu-ray disc image taken by PCM lens with 38 μm BTG microsphere embedded at different WD. Scale bars in (a) is 1μm and in (b)-(i) are 10μm.

## 3.2 Scanning imaging

Above results have verified the imaging performance of developed unibody PCM-objective where the super resolving ability is mainly determined by the PCM lens within certain WD range. Even though, the FOV for super-resolution is limited to the central part of microsphere. To expand the viewable area, therefore, scanning imaging is necessary. In our unibody PCM-objective, the PCM lens is synchronized with an objective lens. During the scanning operation, it is beneficial to keep relative position of both parts of the lens being constant. Meanwhile, different from our previous work [22], the unibody PCM-objective operates in air, it is now friction-free and no lubricant media between lens and object is needed. Due to the elimination of dragging issue, we can easily and precisely move sample to the designated location with knowing its relative coordinate. Therefore, image stitching becomes easier. In this study, scanning imaging was performed with a well-structured semiconductor IC chip and Blu-ray disc sample to demonstrate the feasibility of scanning imaging using a developed unibody PCM-objective, which was shown in Fig. 4. The IC chip in Fig. 4(a) same as used in Fig. 2, which has features of 200 nm and 400 nm, was imaged by 38 μm microsphere with WD around 2.8 μm. Then the sample was moved in XY plane and kept constant in Z direction to get same quality image in every location. The scanning path followed line-by-line scanning, as yellow dash line showing in Fig. 4(b). The camera captured one frame for every 1.2 μm the stage travelled and totally 10×10 frames were recorded. Large-area imaging can be achieved by stitching together those 100 recorded frames, and the finalized image was demonstrated in Fig. 4(b). Similar scanning experiments were also performed to a Blu-ray disc sample [Fig. 4(c, d)] and more complex structure on IC chip [Fig. 4(e, f)]. The scanning processes were video recorded (Visualization 1-3).

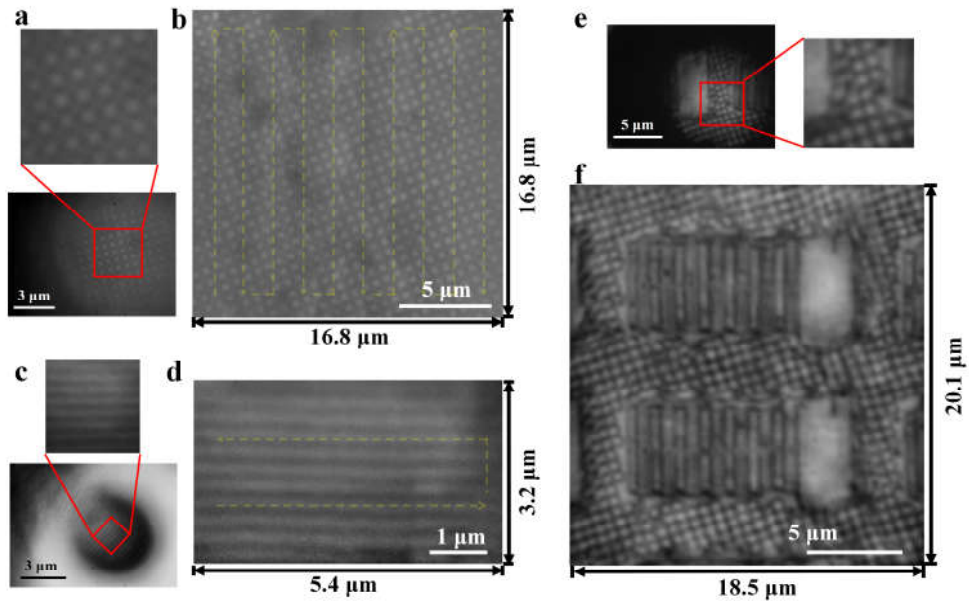

Fig. 4. Scanning super-resolution imaging and image stitching. (a) Single frame image of IC chip sample. (b) Stitched image of 10 x 10 frames. (c) Single frame image of Blu-ray disc. (d) Stitched image of 20 x 2 frames. (e) Single frame image of complex structure of IC chip. (f) Stitched image of 13 x 10 frames.

### 3.3 Theoretical analysis

In the experiment, we found that the encapsulated status of microsphere by the PDMS might be slightly varied due to microsphere size, the thickness of PDMS layer and process deviation. To further investigate whether this deviation would affect imaging performance and to understand focusing characteristics of PCM lens in different capped cases, computational modeling was performed using in-house developed code based on physical optics. Since the curvature radius of the Plano-convex lens is 15.5 mm, which is much greater than the size of the microsphere, the curved surface can be regarded as a plane surface in the simulation range. A planar wave (470 nm) with x-polarization was set to propagate through a 38 μm BTG microsphere ($n_p$=1.9) partially encapsulated by a PDMS layer ($n_m$=1.4) to imitate PCM lens in the experiment.

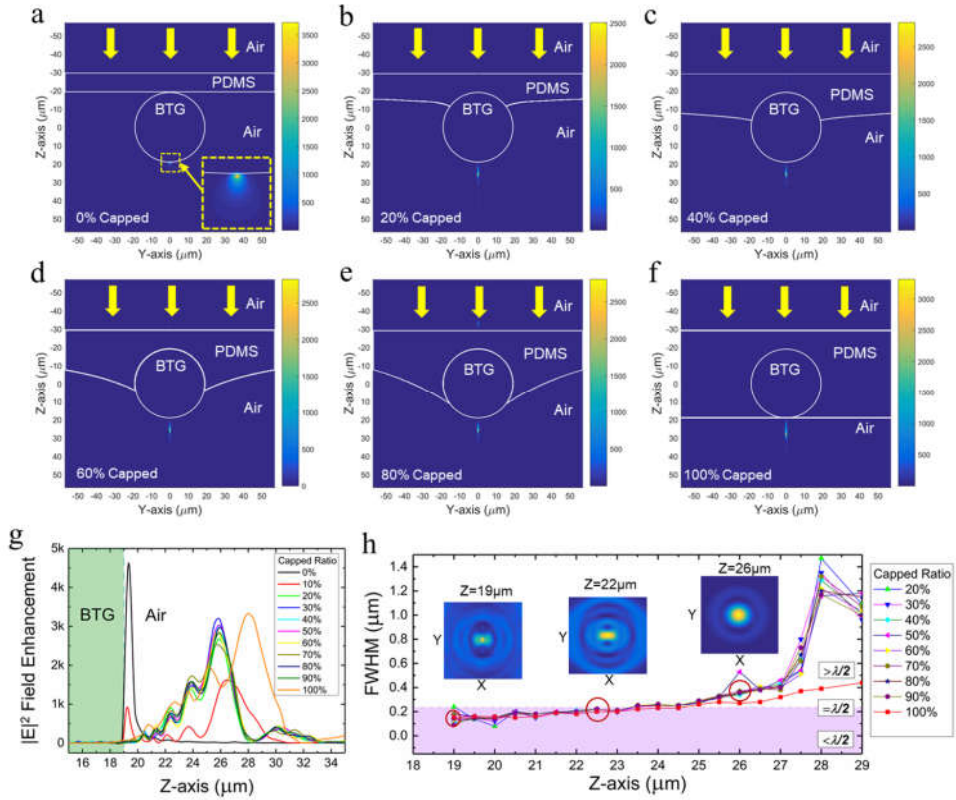

Fig. 5. Modelling of PCM lens focusing properties. YZ-plane electric field intensity distribution of 38 μm BTG microsphere encapsulated by PDMS with (a) 0%, (b) 20%, (c) 40%, (d) 60%, (e) 80%, (f) 100%, respectively. (g) Comparison of $|E|^2$ enhancement along z-axis between different capsulation cases. (h) Comparison of FWHM along y-axis between different capsulation cases.

Here, comparisons were made between different capsulation conditions, including PDMS immersion ratio from 0% to 100%. Fig. 5(a-f) show the calculated optical field distribution in YZ-plane in the respective capsulation cases. In case of non-immersed BTG microsphere [Fig. 5(a)], the planar wave is focused at boundary of the shadow side and then quickly diverged. Under this circumstance, although field enhancement can reach over 3500×, magnified virtual image cannot be generated. When the BTG microsphere begins to submerge in the PDMS, its focus moves into the air region. Fig. 5(b-f) reveal that the light wave is converged after passing through capped particle lenses and focused hot spots exhibit elongated shape and then diverge with distance. The maximum intensity is found several microns distance away from the microsphere and quickly drops afterward. As expected, Fig. 5(f) (fully-capped case) shows slightly longer focal spot in the z-direction than other cases. Fig. 5(g) specifically compare intensity profiles along the z-direction of eleven cases. As observed, differently capped microspheres can immensely enhance light by more than 2500 times and, meantime, 20-90% capped microspheres generate very similar intensity distribution along the z-axis. Their highest peaks appear at around $z=26$ μm (WD=7μm) in air. This reveals that varied formations of encapsulation in range of 20-90% have a negligible effect on their focal length. In addition, capsulations of 0-10% produce completely different focusing, either single peak occurring at the boundary. Furthermore, the calculated spot size, full-width at half-maximum (FWHM) along the y-direction, corresponding to the WD is illustrated in Fig. 5(h). Here, we only consider the capsulation case which can generate virtual image and 0-10% capped cases were not considered. In 20-100% capsulations, we discovered that the variations of spot size in different capped modes basically follow the same trend: the FWHM fluctuates slightly between 0.2-0.4 μm and grows slowly until the focus point (z=26μm), then jumps to micro-scale due to light divergence. It is noted that when distance less than 23.5 μm (WD<4.5 μm) the spot sizes in all cases are slightly better than the diffraction limit ($\lambda/2=0.235$ μm), i.e., a weak super-resolution was achieved in air. Outside this range, the sub-diffraction focusing is lost. Moreover, we picked three points in 80% capped condition to plot lateral focusing profile. As three inserts presented in Fig. 5(h), the central spot shapes at 19 and 22 μm exhibit non-circular pattern but smaller width in y-direction and some peripheral concentric rings patterns are found. While a circular hot spot is generated at focal point (z=26μm) and surrounding rings patterns are suppressed. As previous experimental results described, imaging performance is affected by the WD and super-resolution phenomenon happens within ~4 μm WD. This is thus consistent with the presented theoretical simulations. We may draw conclusions from the above simulation results that the formations of microsphere capsulation in range of 20-90% have an insignificant effect on focusing performance, including focal length, intensity enhancement, and imaging resolution. This is beneficially for manufacturing PCM lens because subtle process deviation would not result in an obvious discrepancy in imaging performance.

The present PCM-objective, with ability to resolve 100 nm features in a far-field setup in the air, is a "modest performance" super-resolution objective lens which doesn't deliver as strong super-resolution as other forms of near-field dielectric superlenses (e.g. $TiO_2$ [21], hybrid nano-lens [31]). It is still, however, very unique, and has clear advantages of performing reliable scanning super-resolution imaging in dry condition. The design holds great promise for commercial developments. A higher resolution PCM-objective, with the ability to resolve 50-70 nm features in the air with micrometer scale WD, would be achieved by using smaller-sized BTG microsphere (diameter <= 20 μm) or high-index $TiO_2$ microsphere (n=2.4) owing to their stronger light confinement effects. This is currently in progress. A new physical effect – optical super resonance in microsphere due to high-order Fano resonance – could be another new route to achieve extreme super-resolution in PCM-objective and other microsphere-based techniques but will require a new modulated lighting source to match the super-resonance conditions [32].

## 4. Conclusion

In summary, we have developed a unibody PCM-objective by tactfully integrating microsphere lens onto an ordinary microscope objective lens. It is advantageous to simplify the imaging system and improve usability in practical operation. The resolution of the developed lens in different WD was experimentally and theoretically investigated. As results indicated, 100 nm Blu-ray disc features were discernible in air condition by a 38μm microsphere PCM lens within 4 μm WD. Such WD provides a possibility for non-invasive scanning imaging. This lens design is versatile, making it convenient to turn a normal microscope into a super-resolution nanoscope without affording too much expense. Moreover, it has potential to other super-resolution applications including sensing, trapping, laser marking and manipulation of nano-objects and samples.